\documentclass[a4paper,11pt]{article}
\usepackage{pos}
\usepackage{graphicx}
\usepackage{ulem}
\usepackage{color}
\usepackage{xcolor}
\usepackage{url}

\usepackage{amsmath,amssymb}
\usepackage{longtable}
\usepackage{caption}
\usepackage{subcaption}
\usepackage{slashed}
\usepackage{cleveref}
\usepackage{physics}
\usepackage{acronym}
\AtBeginDocument{%
  \let\oldac\ac
  \renewcommand\ac[1]{\leavevmode\oldac{#1}}%
  \expandafter\def\csname AC@verridelabel\endcsname#1{}%
}
\usepackage{dsfont}
\usepackage{float}
\usepackage{comment}

\crefname{section}{sec.}{secs.}
\crefname{table}{tab.}{tabs.}
\crefname{figure}{fig.}{figs.}
\crefname{equation}{eq.}{eqs.}
\crefname{appendix}{appendix}{appendices}

\newcommand{\gV}{g_{V\pi\pi}}
 
\newcommand{\SO}{\mathrm{SO}}
\newcommand{\SU}{\mathrm{SU}}
\newcommand{\U}{\mathrm{U}}
\newcommand{\Sp}{\mathrm{Sp}}

\def\lsim{\mathrel{\rlap{\lower4pt\hbox{\hskip1pt$\sim$}}
    \raise1pt\hbox{$<$}}}         
\def\gsim{\mathrel{\rlap{\lower4pt\hbox{\hskip1pt$\sim$}}
    \raise1pt\hbox{$>$}}}

\acrodef{CH}{Composite Higgs}
\acrodef{QCD}{quantum chromodynamics}
\acrodef{pNGB}{pseudo Nambu--Goldstone Boson}
\acrodefplural{pNGB}[pNGBs]{pseudo Nambu--Goldstone Bosons}
\acrodef{SM}{Standard Model}
\acrodef{EW}{electroweak}
\acrodef{irrep}{irreducible representation}
\acrodefplural{irrep}[irreps]{irreducible representations}
\acrodef{VEV}{vacuum expectation value}
\acrodef{UFO}{Universal FeynRules Output}
\acrodef{PC}{partial compositeness}

\title{New massive resonances at the LHC}
\author{Rosy~Caliri}

\affiliation{Institut f\"{u}r Theoretische Physik und Astrophysik, Uni W\"{u}rzburg, Emil-Hilb-Weg 22, D-97074 W\"{u}rzburg, Germany}

\emailAdd{rosy.caliri@uni-wuerzburg.de}

\abstract{ 
We investigate composite Higgs models arising from a gauge-fermionic UV
completion with partial compositeness, where the Higgs boson emerges as a
pseudo-Nambu-Goldstone boson (pNGB), and top-partners appear as bound
states of three hyperfermions, with $\mathrm{SU}(2)_L \times \mathrm{SU}(2)_R$
as the unbroken global subgroup. A bunch of new bound states is predicted by composite Higgs models: 
spin-0, spin-1/2 and spin-1 resonances. In this work we focus on the phenomenology of the spin-1 resonances. We
demonstrate that a generic prediction of these models is the existence of
two neutral and one charged spin-1 resonances, which mix significantly with
the SM gauge bosons. This mixing allows their single production in
Drell-Yan processes at the LHC. We study the LHC phenomenology of
these states and identify scenarios where their masses could be as low as
$1.5~\mathrm{TeV}$, consistent with the existing LHC data. }

\FullConference{
14th Edition of the Large Hadron Collider Physics (LHCP2026)\\
18-22 May 2026\\
Paris, France
}

\begin{document}
\maketitle
 
\section{Introduction}
\ac{CH} models are strongly interacting gauge theories that address
the naturalness problem of the \ac{SM} by interpreting the Higgs boson as a composite particle \cite{Kaplan:1983fs, Kaplan:1983sm}.
We focus here on models where the Higgs boson emerges as a bound state (\ac{pNGB})
of two hyperfermions $\psi$, representation of a global symmetry $G^{\psi}$
spontaneously broken into a subgroup $H^{\psi}$, that must contain the custodial symmetry $\SU(2)_L \times \SU(2)_R$ to protect the $\rho$ parameter and contain the Higgs bidoublet.
The three minimal cosets fulfilling these constraints
are $\SU(4)/\Sp(4)$, $\SU(5)/\SO(5)$ and $\SU(4)\times \SU(4)/\SU(4)$.\\
To address also the problem of the origin of the heavy top quark mass, these models were
extended to \ac{PC} \cite{Kaplan:1991dc}, postulating the existence of bound states
of 3 hyperfermions, called top partners T,
that mix with the \ac{SM} top quark, generating its mass.
In order to match the quantum numbers of the top quark, another set of hyperfermions 
$\chi$ is introduced, that carry QCD color and hypercharge \cite{Barnard:2013zea,Ferretti:2013kya}.
These hyperfermions are also a representation of a group $G^{\chi}$ spontaneously 
broken into a subgroup $H^{\chi}$, that must contain $\SU(3)_c \times \U(1)_X$. Here, the minimal cosets are 
$\SU(6)/\Sp(6)$, $\SU(6)/\SO(6)$ and $\SU(3)\times \SU(3)/\SU(3)$.\\
This led to the classification of 12 minimal models with two species of hyperfermions, 
specified in \cite{Ferretti:2016upr,Belyaev:2016ftv}.
These models generate a vast new particle spectrum: spin-1/2, spin-0,
and spin-1 resonances in both color and \ac{EW} sectors. The spin-1 resonances are classified as Vector or Axial vectors
depending on whether they decay into two or three \acp{pNGB}.
The phenomenology of these various resonances has been studied in the literature \cite{Ferretti:2016upr,Agugliaro:2018vsu,Cacciapaglia:2022bax,Flacke:2023eil,Cacciapaglia:2019bqz,BuarqueFranzosi:2021kky, Cacciapaglia:2015eqa,Belyaev:2016ftv,Cacciapaglia:2020vyf,Bizot:2018tds,Xie:2019gya,Cacciapaglia:2019zmj,BuarqueFranzosi:2016ooy,Cacciapaglia:2024wdn, Flacke:2026fxb}. 
Here, we will focus on the phenomenology of \ac{EW} spin-1 resonances.
From the model-building point of view, following Refs.~\cite{BuarqueFranzosi:2016ooy,Cacciapaglia:2024wdn}, 
we adopt the hidden gauge symmetry approach \cite{Bando:1987br} which 
introduces a local copy of the global symmetry using the CCWZ construction \cite{Coleman:1969sm,Callan:1969sn}.
For further details we refer to the original paper  \cite{Caliri:2024qjv}.\\
All results shown here are for the $\SU(5)/\SO(5)$ coset,
the other two cosets yield similar conclusions, as discussed in Ref. \cite{Caliri:2024qjv}.

\section{Spin-1 resonances mixing}
As a consequence of the fact that $\SO(5)$ contains $\SU(2)_L \times \SU(2)_R$ as a subgroup, 
some of these new Spin-1 resonances mix with the SM gauge bosons W, Z. If we denote as $r^+= (\tilde W^+_\mu,\, a^+_\mu,\, v_{1\mu}^+,\, v_{2\mu}^+)$, $ r^0= (B_\mu, \tilde W^3_\mu,\, a^0_\mu,\, v_{1\mu}^0,\, v_{2\mu}^0)$  (same as in ref.~\cite{BuarqueFranzosi:2016ooy}) the gauge eigenstates and as $R^+= (W^+_\mu,\, V_{1\mu}^+,\, V_{2\mu}^+,\, V_{3\mu}^+)$, $R^0= (A_\mu, Z_\mu,\, V_{1\mu}^0,\, V_{2\mu}^0,\, V_{3\mu}^0)$  the mass eigenstates, we have
that they are related to each other by mixing matrices $\mathcal{C}$ and $\mathcal{N}$ as follows:
    \begin{center}
        $ 
        \begin{array}{cc}
            r^+ = \mathcal{C} R^+ 
             & 
            r^0= \mathcal{N} R^0
        \end{array}
      $       
     
    \end{center}

A generic prediction of these models is the existence of two neutral $V_1^0, \;V_2^0 $
 and one charged $V_1^+ $ spin-1 resonances, which mix significantly with the SM gauge bosons.\\
A consequence of this mixing is that the couplings of the SM gauge bosons to fermions 
get modified and these new spin-1 resonances couple to SM fermions.
For example, in the charged currents:
\begin{equation*}
        \mathcal{L}_{CC} = \frac{\hat{g}}{\sqrt{2}} \sum_{ij} \bar{q}_{i} \gamma^{\mu} \tilde{W}^+_{\mu}  P_L (V_{CKM})^{i,j}  q_{j} = \frac{\hat{g}}{\sqrt{2}} \sum_{i,j, m} \mathcal{C}_{1 m} \bar{q}_{i} \gamma^{\mu} R^+_{m, \mu} P_L (V_{CKM})^{ij} q_{j},
\end{equation*}
 $\tilde{W}^+_{\mu}$ is the gauge eigenstate. The first row of $\mathcal{C}$,
that rotates the gauge eigenstates into the mass eigenstates, gives the couplings of the mass eigenstates to SM fermions.
Similar conclusions can be obtained for the neutral current interactions.
For more details we refer to the original paper \cite{Caliri:2024qjv}. This mixing allows their single production in Drell-Yan processes at the LHC.

\section{Phenomenology}
We focus here on those states which strongly mix with
the SM electroweak bosons: $V^+_1$, $V^0_1$ and $V^0_2$.
The first two states stem essentially from $(3,1)$ of $\SU(2)_L\times \SU(2)_R$
whereas $V^0_2$ is mainly the neutral state of
$(1,3)$ mixing primarily with the hypercharge boson.\\
In view of LHC phenomenology we group the various
decay channels as follows
\begin{alignat}{3}
    &\mathcal V^0 \to q\bar q, \, l^+ l^-, \,\nu\bar \nu, \qquad\qquad &&\mathcal V^0 \to t\bar t, \qquad\qquad &&\mathcal V^0 \to \pi\pi,\, HZ,\, W^+ W^-, \\
    &\mathcal V^+ \to q\bar q', \, l^+ \nu, \qquad &&\mathcal V^+ \to t\bar b, \qquad &&\mathcal V^+\to \pi\pi , \, W^+ Z,\, W^+ H.
\end{alignat}

\begin{figure}[t]
    \centering
    \begin{subfigure}{.32\textwidth}
        \includegraphics[width=\linewidth]{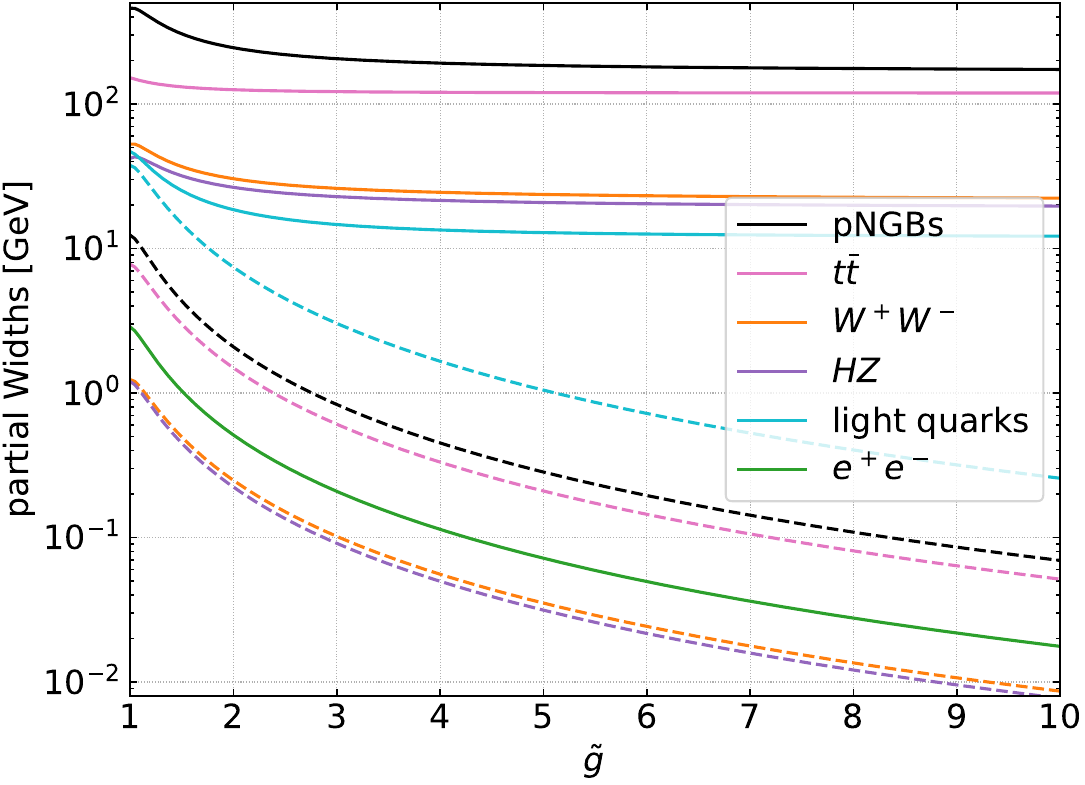}
        \caption{partial widths of $V_{1\mu}^0$}
    \end{subfigure}
    \begin{subfigure}{.32\textwidth}
        \includegraphics[width=\linewidth]{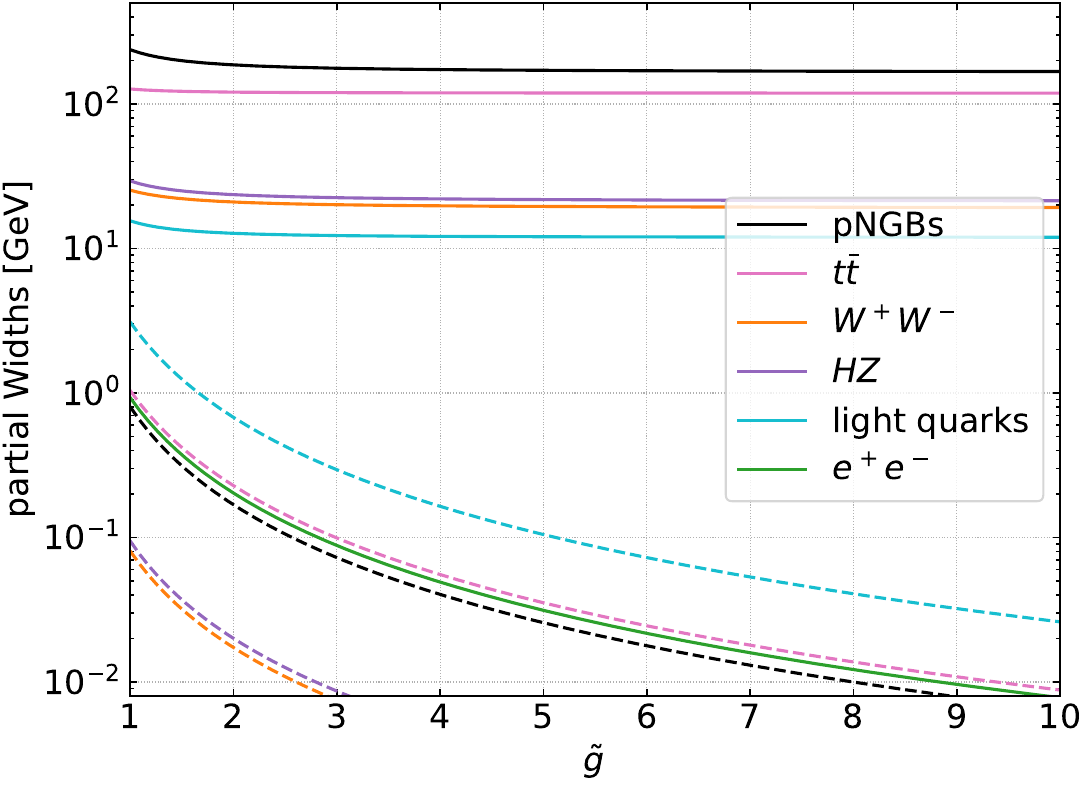}
        \caption{partial widths of $V_{2\mu}^0$}
    \end{subfigure}
    \begin{subfigure}{.32\textwidth}
        \includegraphics[width=\linewidth]{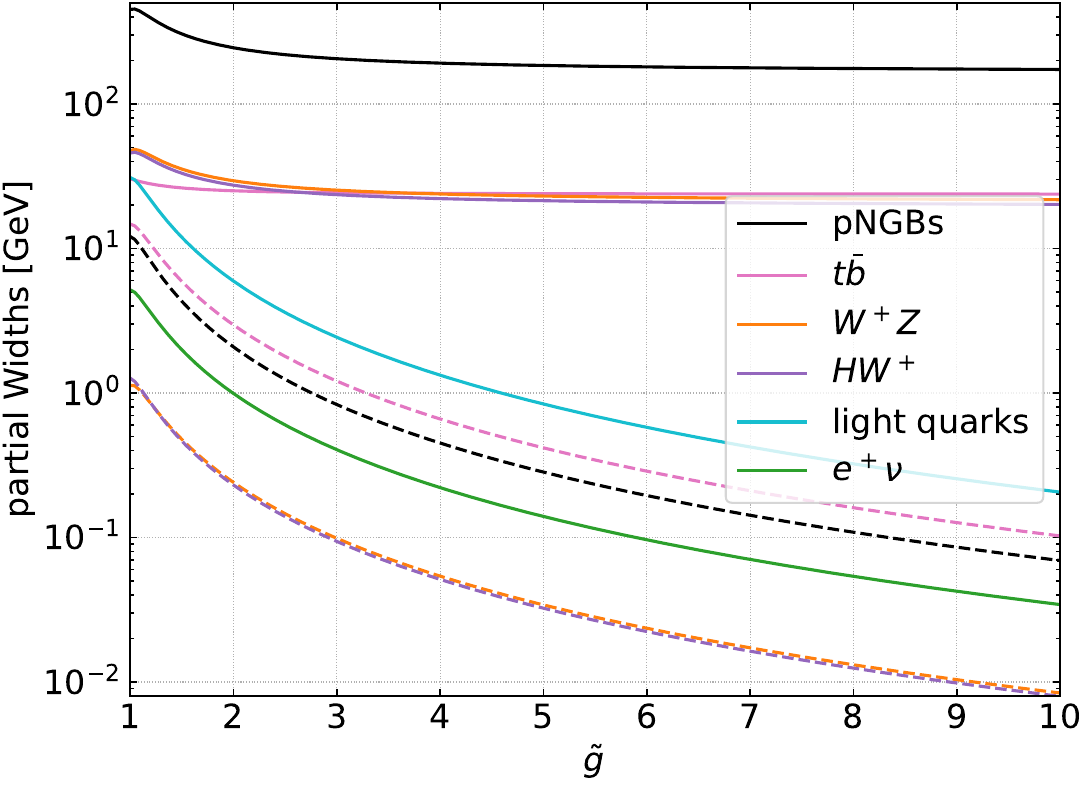}
        \caption{partial widths of $V_{1\mu}^+$}
    \end{subfigure}
    \caption{Partial decay widths. The solid lines of the pNGB, $W^+ W^-$, $HZ$, $W^+Z$ and $H W^+$ channels correspond to a scenario with $\mathbf{strong} \, \boldsymbol \pi$, while the corresponding dashed lines correspond to $\mathbf{weak} \, \boldsymbol \pi$. For the top quark channels, the solid lines correspond to $\mathbf{PC} \, \boldsymbol t$ and the dashed lines to SM-like couplings.
   We have set $M_V=3000$~GeV and $M_\pi=700$~GeV.}
  \label{fig:partial-widths}
\end{figure}

The phenomenology of these resonances depends on various unknown independent
parameters. We choose to parametrize the Lagrangian in terms of: $M_V$ (not the physical mass
but) the mass parameter of the vectors, $\tilde{g}$ the gauge coupling
of the spin-1 resonances, $\xi = \frac{M_A}{M_V}$ the fraction between the mass 
parameter of the axial and vector resonances,
$f_\pi$ the pion decay constant and $g_{V\pi\pi}$ the coupling of the spin-1 resonances to two \acp{pNGB}.\\
In addition, we have the couplings of the spin-1 resonances to the top quark, 
which can be either SM-like or enhanced in case of \ac{PC} where the third generation 
quarks get an additional contribution from the mixing between the elementary fields and the top partners.\\
In particular, we use combinations of the \ac{pNGB} couplings, set to be $\mathbf {weak} \, \boldsymbol \pi$ ($\gV = 0$) or $\mathbf {strong} \,\boldsymbol \pi$ ($\gV = 4$),
and the top quark couplings, fixed to be \ac{SM}-like ($\mathbf{SM} \,\boldsymbol t$) or \ac{PC}-like ($\mathbf{PC}\, \boldsymbol t$) for defining four different scenarios. We expect that a realistic scenario will be in between these extreme cases.\\
To give a better understanding of these scenarios, in \cref{fig:partial-widths}
we show the partial widths of the strongly mixed spin-1 resonances.
Here we can see how going from scenarios with $\mathbf{PC}\, \boldsymbol t$ to 
$\mathbf{SM} \, \boldsymbol t$ the partial width into $t\bar{t}$ decreases, 
instead going from $\mathbf{strong} \, \boldsymbol \pi$ to $\mathbf{weak} \, \boldsymbol \pi$
the partial width into two \acp{pNGB} decreases, same as the decay into
two \ac{SM} gauge bosons or one Higgs and a \ac{SM} gauge boson, due to the longitudinal polarization of the vector boson.\\

In the decays into pNGBs we have to take cascade decays into account 
\cite{Agugliaro:2018vsu,Cacciapaglia:2022bax,Ferretti:2016upr}. 
In case of the $\SU(5)/\SO(5)$ coset,
we need to consider two different scenarios:\\
In the fermiophilic scenario, the \acp{pNGB} dominantly decay into third generation quarks,
\begin{equation}
    S^0 \to t \bar{t}\,,\quad b \bar{b} \qquad
    S^+ \to t \bar{b} \qquad
    S^{++} \to W^+ t \bar{b} \nonumber
\end{equation}
whereas in the fermiophobic scenario, the decays into two SM vector bosons induced by the anomalous WZW terms become relevant \cite{Cacciapaglia:2022bax}. 
\begin{equation*}
    S \to W^+ W^- \gamma,\,W^+ W^-Z , Z Z Z ,\, Z Z \gamma ,\, Z \gamma \gamma
\end{equation*}
We refer to the original paper for further details.\\

\subsection{Constraints from LHC data}
Combining the single production of the vector states with the decay channels 
leads to multiple signatures that have been searched for at the LHC. In particular, 
searches for heavy gauge bosons $W'$ and $Z'$ are relevant 
for us \cite{ATLAS:2019erb,ATLAS:2020lks,ATLAS:2019lsy,ATLAS:2023ibb} to constrain the parameter space of our models
\footnote{We first implemented the Lagrangian in \texttt{FeynRules} \cite{Alloul:2013bka}
format to produce a \ac{UFO} library \cite{Degrande:2011ua}, which was then
loaded into \texttt{MadGraph5\_aMC@NLO} \cite{Alwall:2014hca} v3.5.3 for event generation at $\sqrt{s} = 13$~TeV.
The dynamical renormalization and factorization scales were set via the
\texttt{LHAPDF} \cite{Buckley:2014ana} interface, adopting the \texttt{NNPDF~2.3} set of parton distribution functions \cite{Ball:2012cx}.}. 
For each process we compute the cross sections by scanning a grid of parameter points in the $M_V$--$\tilde g$-plane and compare them to the upper limits obtained from
the above searches to derive exclusion limits. \\
Since no dedicated experimental search covers the decay of spin-1 resonances into 
two bosons, constraints are instead derived through recasting existing searches
\footnote{The generated events are passed through \texttt{Pythia8} \cite{Sjostrand:2014zea} for showering and hadronization, producing a \texttt{HepMC} file \cite{Dobbs:2001ck}.
This is then analyzed with \texttt{MadAnalysis5} \cite{Conte:2012fm,Conte:2014zja,Dumont:2014tja,Conte:2018vmg} v1.10.9beta 
and \texttt{CheckMATE} \cite{Drees:2013wra,Dercks:2016npn} commit number \texttt{1cb3f7}.
These tools cluster jets with the anti-$k_T$ algorithm \cite{Cacciari:2008gp} 
implemented in \texttt{FastJet} \cite{Cacciari:2011ma} 
simulating the detector with \texttt{Delphes 3} \cite{deFavereau:2013fsa}.
The kinematic cuts are applied to the events, 
and the number of remaining events is used to calculate an exclusion value 
using the CL$_s$ method \cite{Read:2002hq} for each signal region.
For each search and signal region, the observed exclusion with
 the strongest expected bound is then collected. In addition, the events are confronted with the \ac{SM} with
\texttt{Rivet} \cite{Bierlich:2019rhm} v3.1.8, and
\texttt{Contur} \cite{Butterworth:2019wnt,Buckley:2021neu} v2.4.4.}.
The final reported result corresponds to the strongest exclusion obtained from any single search;
no statistical combination is performed beyond what is implemented in the tools.
The 95\% CL exclusion contour is then drawn in the $M_V$--$\tilde{g}$ plane.

\begin{figure}[t]
    \centering
    \includegraphics[width=0.5\linewidth]{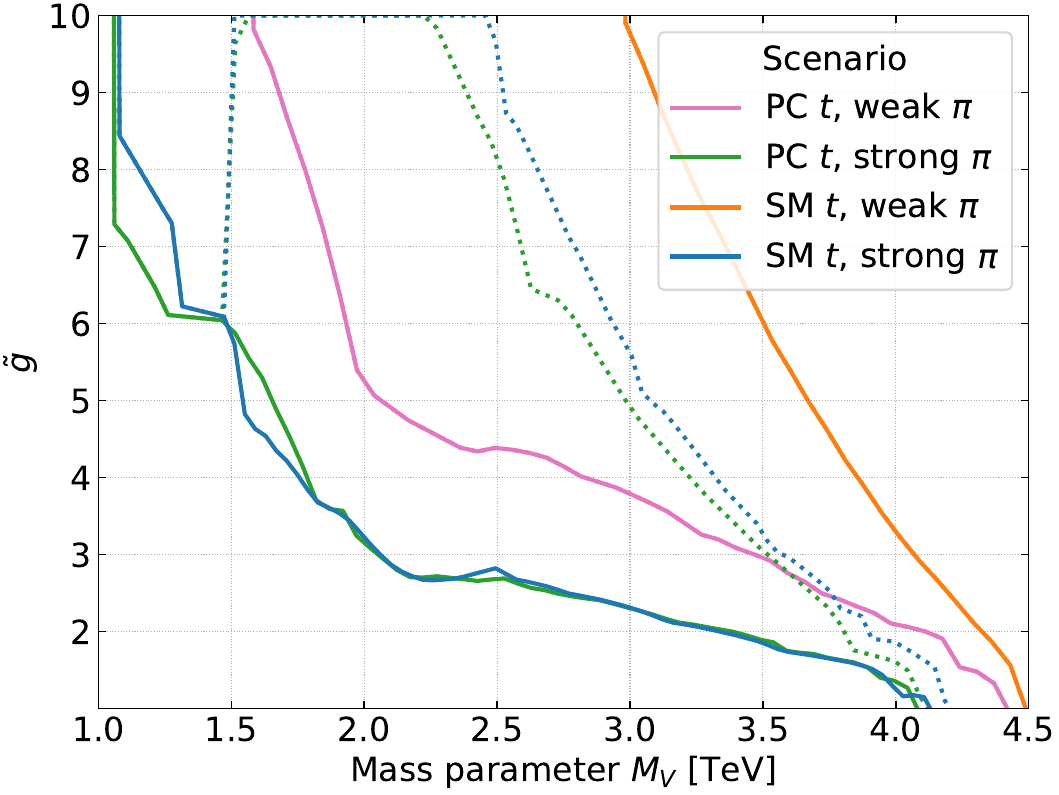}
    \caption{Bounds on the single production of heavy vectors. For each scenario we show the envelope of the bounds from the individual channels. The solid lines correspond to the fermiophilic, the dotted lines to the fermiophobic model, both with $M_\pi = 700\,\mathrm{GeV}$.}
    \label{fig:envelopes_su5so5}
\end{figure}

The resulting bounds for each channel are shown in Ref. \cite{Caliri:2024qjv}. Here, instead, we 
show the combined bounds for all scenarios (\cref{fig:envelopes_su5so5}) to 
find which regions of the parameter space are still viable. We show the fermiophilic 
scenario as a solid line and the fermiophobic one with a dotted line.
The scenario with weak couplings to top and \acp{pNGB} (orange) is strongly constrained to $M_V > 3$~TeV to 4.5~TeV.
If only the \ac{PC} couplings are turned on (pink), the bounds are substantially weaker,
for moderate values of $\tilde g$, $M_V$ as low as 2~TeV remains viable. 
The parameter space is further allowed if we also include $\mathbf{strong} \, \boldsymbol \pi$ (green).
The fermiophobic scenario is more strongly constrained than the fermiophilic one, 
with the latter allowing $\tilde g>4$ for $M_V > 2$~TeV. There are not too many differences in the scenario with SM couplings to the top and a strong coupling to the \ac{pNGB}s (blue).
Overall, the case with large $\gV$ couplings leaves the 
largest portion of parameter space open, particularly in the fermiophilic case.

The results so far have been for the case of the
$\SU(5)/\SO(5)$ coset. The other two
cosets differ mainly in the \ac{pNGB} sector.
The overall results are nevertheless very similar
to the previous coset as can be seen in the paper \cite{Caliri:2024qjv}.

\section{Conclusions}
\label{sec:outlook}

We have studied the LHC phenomenology of electroweak spin-1 resonances in \ac{CH} 
models with fermionic UV completions in the minimal cosets \cite{Ferretti:2016upr,Belyaev:2016ftv}.
We focused on those states which can mix
with the electroweak vector bosons of the SM, so they can be singly produced at the LHC. We have found
that independent of the coset there is always one charged spin-1 resonance mixing
sizably with
the W-boson and two neutral spin-1 resonances with the Z-boson. We have derived bounds
in the mass-coupling plane for all cosets both from direct searches, when possible, and from recast searches otherwise.
We have considered four different
scenarios to study the effect of unknown model-dependent couplings.
Masses as low as about 1.5 TeV are still viable by current LHC data.
In such scenarios, also the states which
only mix weakly or not at all will have masses
of about 1.5 TeV.

\section*{Acknowledgements}
I thank 
Jan Hadlik, Manuel Kunkel, Werner Porod and Christian Verollet for collaboration on this project. This work has been supported by DFG, project 
no.~PO-1337/12-1. This work was supported by the German Academic Exchange Service
(DAAD), PROCOPE project nr. 57755908.
\clearpage
\bibliographystyle{utphys}
\bibliography{main}

\end{document}